%% file: dark_matter.tex
\documentclass[preprint,12pt]{elsarticle}

\usepackage{graphicx}

\usepackage{amssymb}
\usepackage{subfig}
\biboptions{sort&compress}

\input{config}
\journal{Physics Letters B}

\begin{document}

\begin{frontmatter}


%
\title{The impact of a 126 GeV Higgs on the neutralino mass}

\author[label1]{C. Beskidt}\ead{conny.beskidt@kit.edu}
\author[label1]{W. de Boer}\ead{wim.de.boer@kit.edu}
\author[label1,label2]{D.I. Kazakov}
\address[label1]{Institut f\"ur Experimentelle Kernphysik, Karlsruhe Institute of Technology, P.O. Box 6980, 76128 Karlsruhe, Germany}
\address[label2]{Bogoliubov Laboratory of Theoretical Physics, Joint Institute for Nuclear Research, 141980, 6 Joliot-Curie, Dubna, Moscow Region, Russia}

\begin{abstract}
We highlight the differences of the dark matter sector between the constrained minimal supersymmetric SM (CMSSM) and the next-to-minimal supersymmetric SM (NMSSM) including the 126 GeV Higgs boson using GUT scale parameters. In the dark matter sector the two models are quite ortho\-gonal: in the CMSSM the WIMP is largely a bino and requires large masses from the LHC constraints. In the NMSSM the WIMP has a large singlino component and is therefore independent of the LHC SUSY mass limits. The light NMSSM neutralino mass range is of interest for the hints concerning light WIMPs in the Fermi data. Such low mass WIMPs cannot be explained in the CMSSM. Furthermore, prospects for discovery of XENON1T and LHC at 14 TeV are given.

\end{abstract}

\end{frontmatter}


\section{Introduction}
\label{Introduction}

Within Supersymmetry (SUSY)\cite{Martin:1997ns,Haber:1984rc,deBoer:1994dg}, a light Higgs boson below 135 GeV is predicted, so the discovery of a Higgs-like boson with a mass of 126 GeV \cite{Aad:2012tfa,Chatrchyan:2012ufa} strongly supports SUSY despite  the fact that no SUSY particle has been found so far. However, the precise value of the Higgs mass depends on radiative corrections. To include the radiative corrections between the GUT scale and the electroweak scale we use the constrained supersymmetric models, which assume unification of the gauge couplings and unification of the SUSY masses at the GUT scale. Within the constrained minimal supersymmetric standard model (CMSSM) \cite{Kane:1993td} a 126 GeV Higgs boson is only possible for stop masses above 1 TeV, see e.g. \cite{Beskidt:2012sk,Buchmueller:2013rsa,Fowlie:2012im,Bechtle:2013mda} and references therein. However, a 126 GeV Higgs boson is easily obtained in the semi-constrained next-to-minimal supersymmetric standard model (NMSSM) \cite{Ellwanger:2009dp} for small and moderate stop masses, because the mixing with the additional Higgs singlet increases the Higgs mass at tree level \cite{King:2012tr,Cao:2012fz,Belanger:2012tt,Ellwanger:2012ke,King:2012is,Vasquez:2012hn,Beskidt:2013gia,Badziak:2013bda}. The additional terms at tree level are only significant if both new couplings in the NMSSM Higgs sector $\lambda$ and $\kappa$ are significant. This is not possible in the constrained NMSSM (CNMSSM), which favors small values for $\lambda \rightarrow 0$, see e.g. Ref. \cite{Kowalska:2012gs}. Furthermore $\kappa$ is not an independent free parameter in the CNMSSM and becomes small as well. In this case the CNMSSM and CMSSM are degenerate for the SUSY sector. In the semi-constrained NMSSM the couplings $\lambda$ and $\kappa$ are independent and free parameters. The semi-constrained NMSSM implies non-universal masses of the neutral Higgs doublets at the GUT scale, as it is the case in the non-universal Higgs model (NUHM), which can be considered an extension of the CMSSM with non-universal Higgs masses. However, since the NUHM and CMSSM are only different in a small region of parameter space \cite{Buchmueller:2013rsa}, we only consider the CMSSM.

In both, the CMSSM and NMSSM, the lightest neutralino has all the properties of dark matter (DM) particles \cite{Jungman:1995df}, if DM is made up of WIMPs (Weakly Interacting Massive Particles). Compared to the CMSSM the lightest neutralino has a strong singlino component in the NMSSM, which changes its properties. Especially low WIMP masses are now allowed in contrast to the CMSSM, where its mass is related to other SUSY masses and LHC limits on gluinos require WIMP masses above 180 GeV. So hints for 30 GeV WIMP masses in the Fermi data \cite{Daylan:2014rsa} would find no explanation in the CMSSM, but could be allowed in the NMSSM. 

In this Letter we study the differences in the neutralino sector (e.g. mass ranges and scattering cross sections) considering the allowed parameter space of both models. After a short summary of the neutralino sector in the CMSSM and NMSSM we discuss  global fits to all available data for both models. We conclude by giving future prospects for the discovery reach from direct dark matter searches with XENON1T and the LHC at 14 TeV.

\section{Neutralino sector of the CMSSM and NMSSM}

The NMSSM distinguishes itself from the CMSSM by an additional Higgs singlet in addition to the usual two doublets. The superpartner of the Higgs singlet, the singlino, mixes with the gauginos and Higgsinos.
The resulting mixing matrix reads \cite{Ellwanger:2009dp,Staub:2010ty}:

\beq\label{eq1}
{\cal M}_0 =
\left( \ba{ccccc}
M_1 & 0 & -\frac{g_1 v_d}{\sqrt{2}} & \frac{g_1 v_u}{\sqrt{2}} & 0 \\
0 & M_2 & \frac{g_2 v_d}{\sqrt{2}} & -\frac{g_2 v_u}{\sqrt{2}} & 0 \\
-\frac{g_1 v_d}{\sqrt{2}} & \frac{g_2 v_d}{\sqrt{2}} & 0 & -\mu_\mathrm{eff} & -\lambda v_u \\
\frac{g_1 v_u}{\sqrt{2}} & -\frac{g_2 v_u}{\sqrt{2}}& -\mu_\mathrm{eff}& 0 & -\lambda v_d \\
0& 0& -\lambda v_u&  -\lambda v_d & 2 \kappa s + \mu'
\ea \right)
\eeq

where $M_1$ and $M_2$ are the gaugino masses of the $SU(2) \times U(1)$ group with the gauge couplings $g_1$, $g_2$ and the Higgs mixing parameter $\mu_{eff}$. The couplings $\lambda$ and $\kappa$ describe the coupling with the Higgs singlet s: $\lambda$ is the coupling of the Higgs doublets with the singlet and $\kappa$ describes the singlet self interaction. Furthermore, the vacuum expectation values of the two Higgs doublets $v_d$,$v_u$ and the singlet s enter the neutralino mass matrix. 
The term $\mu'$ in the last diagonal element appears in the general NMSSM only and is set to zero for this analysis. The first $4 \times 4$ elements of the neutralino mixing matrix correspond to the MSSM neutralino mass matrix, see e.g. Ref. \cite{Martin:1997ns}. To obtain the mass eigenstates the mass matrices have to be diagonalized. Typically the diagonal elements in Eq. \ref{eq1} dominate over the off-diagonal terms, so the neutralino masses are of the order of $M_1$, $M_2$, the Higgs mixing parameter $\mu_{eff}$ and in case of the NMSSM $2\kappa s $. 

\begin{table}
\centering
\begin{tabular}{clll}
\hline\noalign{\smallskip}
No. &Constraint & Data & Refs.  \\
\noalign{\smallskip}\hline\noalign{\smallskip}
1& $\bsg$ & $(3.55 \pm 0.24)\cdot 10^{-4}$ & \cite{hfag} \\
 2&  $\btaunu$ &  $(1.68 \pm 0.31)\cdot 10^{-4}$ & \cite{hfag} \\
  3&  $\Delta a_\mu$ & $(302~\pm~63 (exp)~\pm~61 (theo))\cdot 10^{-11}$ &  \cite{Bennett:2006fi}\\
  4& $\bsmm$ &  $ (2.9 \pm 1.1) \cdot 10^{-9}$ & \cite{Aaij:2013aka,Chatrchyan:2013bka}\\
5&$m_A$ & $m_A > 480$ GeV for $\tb \approx 50$& \cite{Chatrchyan:2012vp,Aad:2011rv}\\
6&LEP & $ \xi^2 < 0.003 - 1$& \cite{Barate:2003sz}\\
7&$\Omega h^2$ & $0.1199 \pm 0.002$ &  \cite{Ade:2013lta} \\
8&$m_h$  & $(126 \pm 2)$ GeV & \cite{Aad:2012tfa,Chatrchyan:2012ufa}\\
9&ATLAS & $ \sigma^{SUSY}_{had} < 0.001$ pb & \cite{ATLAS-CONF-2013-047,ATLAS-CONF-2013-061}\\
10&LUX & $\sigma_{\chi N} < 1 \cdot 10^{-9} - 5\cdot 10^{-9} $pb& \cite{Akerib:2013tjd}\\
\noalign{\smallskip}\hline
\end{tabular}
\caption{List of all constraints used in the fit to determine the excluded region of the CMSSM and NMSSM parameter space. }
\label{t1}
\end{table}

\section{Results for the neutralino sector from the global fits}

\begin{figure}[t!] 
\hspace{0.08\textwidth}\textbf{\textsf{CMSSM \hspace{0.35\textwidth} NMSSM}} 
 \centering
\subfloat[]{\label{f1a}\includegraphics[width=0.48\textwidth]{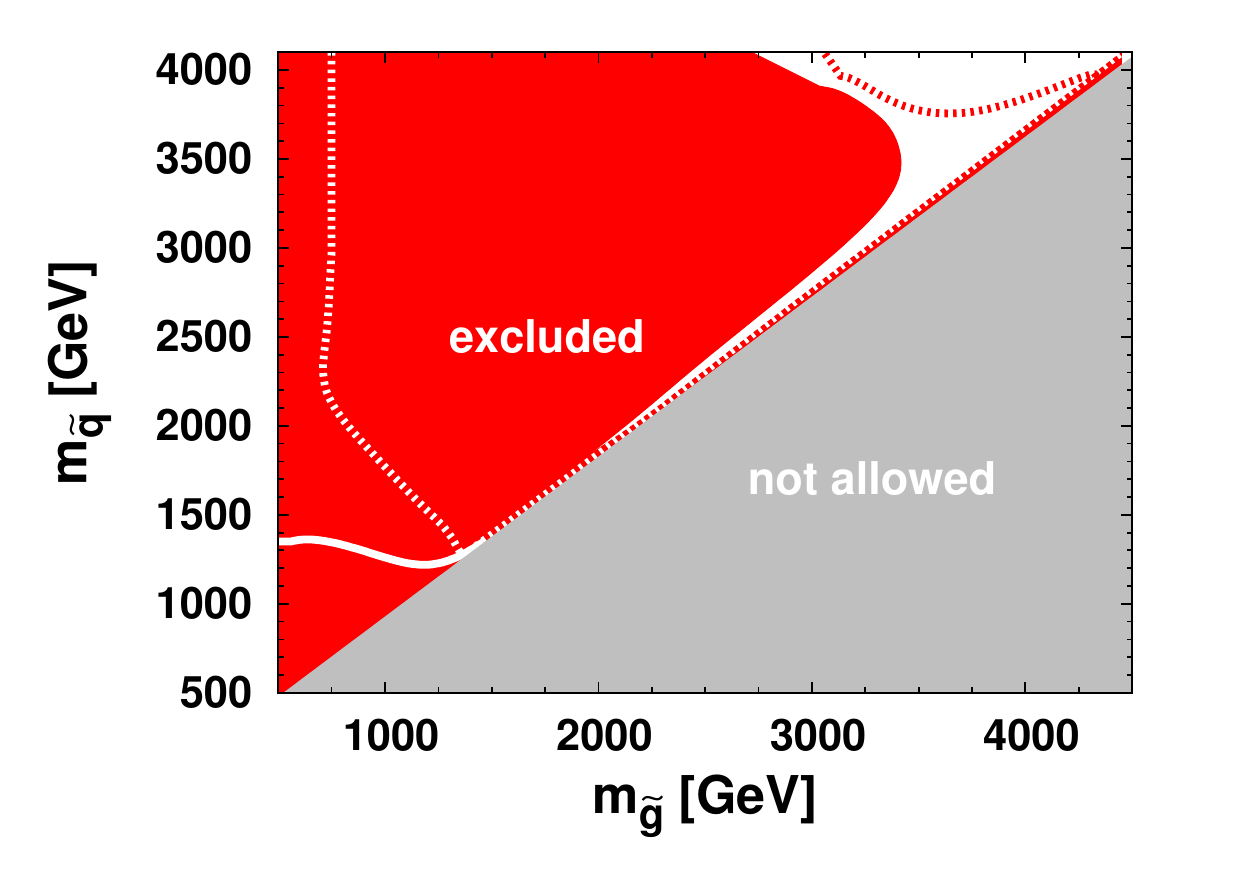}}
\subfloat[]{\label{f1b}\includegraphics[width=0.48\textwidth]{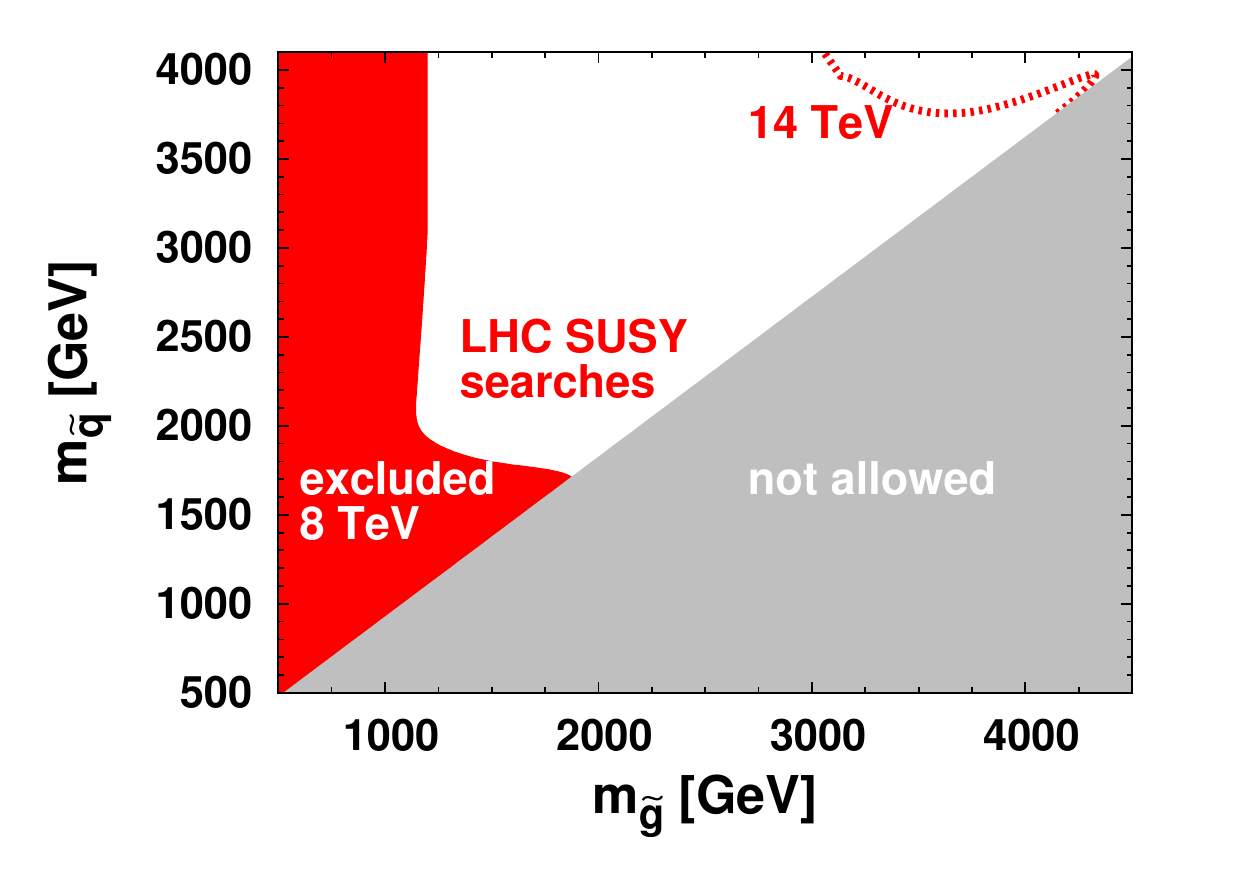}}
 \caption{ (a): The CMSSM excluded region at 95\% C.L. from all constraints in Table \ref{t1} in the squark-gluino mass plane. For the squark mass the averaged value of the two first generations was used. The solid (white) line in the left bottom corner corresponds to the 95\% C.L. exclusion contour obtained by only requiring a Higgs boson of 126 GeV and the dotted (white) line by the combination of a Higgs boson and the relic density constraint. (b): The NMSSM excluded region at 95\% C.L. from  LHC SUSY searches at 8 TeV and 20.1 $fb^{-1}$. The other constraints from Table \ref{t1} do not influence the excluded region. The extrapolation of these searches to 14 TeV and 3000 $fb^{-1}$ is represented by the dotted (red) lines in the top corners. The light (grey) regions are not allowed in constrained models.
}\label{f1}
 \end{figure}

To determine the allowed values of all parameters we perform global fits to all available data given in Table \ref{t1}. All NMSSM/CMSSM observables have been calculated with the publicly available software package  NMSSMTools\_4.1.1 \cite{Das:2011dg}, which has an interface to the micrOMEGAs\_3.6.7 package for calculating the relic density and the WIMP cross sections \cite{Belanger:2013oya}. We apply our multi-step fitting method to cope with the strong correlations of the CMSSM and NMSSM parameters. Details can be found in Refs. \cite{Beskidt:2012sk} and \cite{Beskidt:2013gia}. In the first step we fix the common masses for the spin 0 and spin 1/2 particles at the GUT scale (\mzero~and \mhalf ) and then perform the fits for all possible pairs of \mzero-\mhalf~in the range between 100 GeV and 3(1.5) TeV for the CMSSM(NMSSM). The fits minimize the $\chi^2$ function to restrict the remaining parameters (2 in the CMSSM, 7 in the NMSSM). The two main constraints are coming from the limit on the gluino and squark masses of the order of 1 TeV and the Higgs mass of 126 GeV in combination with the first seven constraints from Table \ref{t1}. The exclusion contours can either be given in the \mzero-\mhalf~plane (see  e.g. Ref. \cite{Beskidt:2013gia}) or  in the squark-gluino mass plane, as shown in Fig. \ref{f1}.

A Higgs mass of 126 GeV\footnote{Recent results from the summer conferences present an averaged mass for the Higgs boson of about 125.2 GeV. The down-shift of the mass by approximately 1 GeV can be easily compensated by slightly different values of the free parameters, which leads to the same conclusions.} requires squark masses above 1200 GeV, as shown by the solid (white) line in Fig. \ref{f1a}. These are the squark masses of the first and second generation. The third generation squarks are usually lighter. The lightest one depends on the splitting between the stops. However, this splitting is restricted by the other constraints especially \bsmm~is important here \cite{Beskidt:2011qf}. Combined with the other constraints the stop masses are typically 200 GeV below the masses of the first and second generation. If one requires in addition the relic density to correspond to the lightest neutralino relic density one obtains the dotted (white) line. If the other constraints of Table \ref{t1} are required as well, one obtains the dark (red) region. Here the \bsmm~constraint requires heavy SUSY masses since the CMSSM requires $\tb \approx 50$ from the relic density constraint \cite{Beskidt:2010va} and $\bsmm \approx \tan^6\beta$, so a strong suppression by heavy, almost degenerate stop masses is needed. The light grey region is not allowed in both constrained models since from the radiative corrections the gluinos have to be heavier than the squarks.  

The dark (red) region in Fig. \ref{f1b} corresponds to the 95\% C.L. exclusion region in the NMSSM, which originates mainly from the LHC SUSY searches at a center-of-mass energy of 8 TeV and an integrated luminosity of 20.1 $fb^{-1}$. Other constraints  of Table \ref{t1} do not play a role, since stop masses well below 1 TeV are allowed and B-physics constraints are automatically fulfilled because of small \tb~values. 
The dotted (red) line in the top right corners represent the extrapolation of the SUSY searches to 14 TeV and 3000 $fb^{-1}$, which will be discussed in more detail in section \ref{future}. 
In the following, we will concentrate on the neutralino masses in the allowed region of parameter space.

\begin{figure}[t!] 
 \centering
\subfloat[]{\label{f2a}\includegraphics[width=0.49\textwidth]{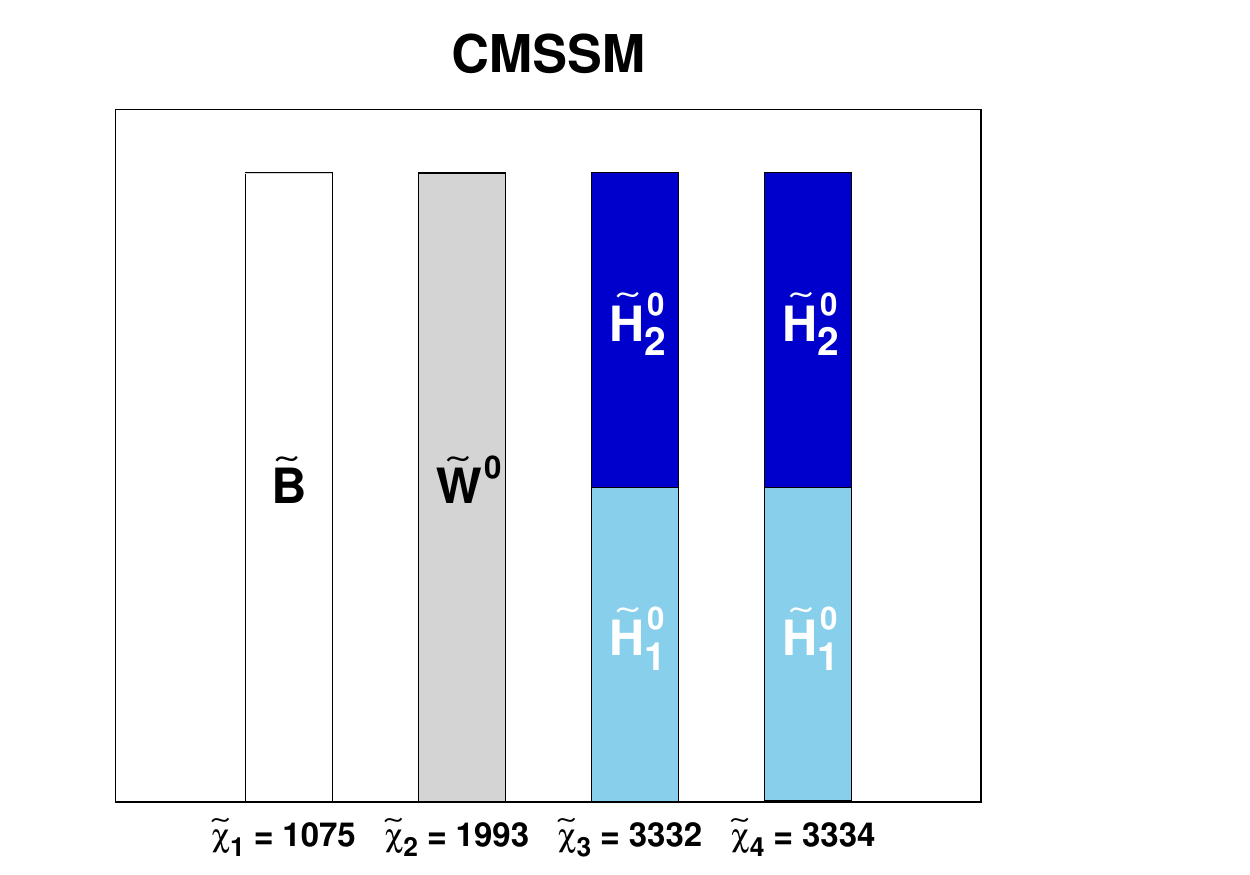}}
\subfloat[]{\label{f2b}\includegraphics[width=0.49\textwidth]{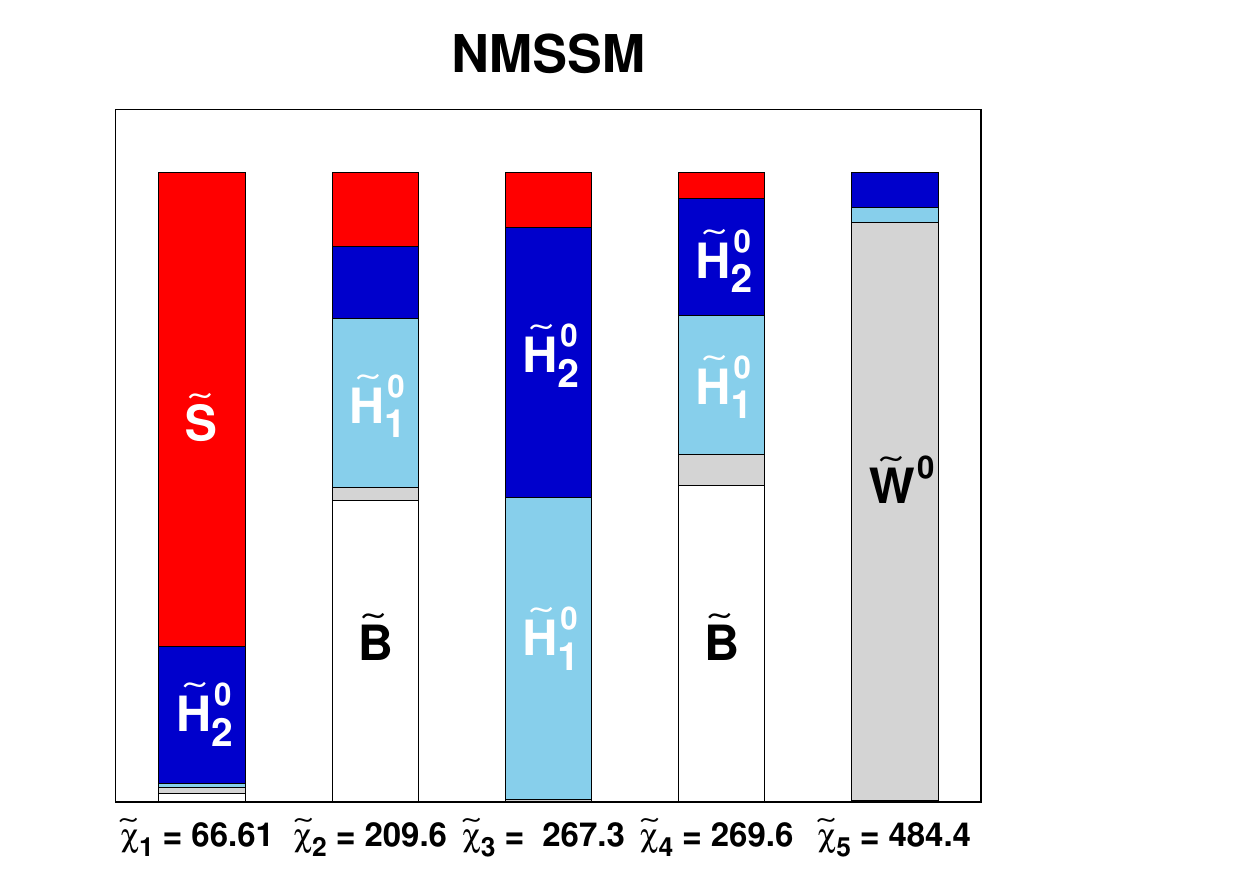}}
 \caption{  The neutralino content for typical points in parameter space for the CMSSM  (a) and NMSSM (b). The neutralino mixing matrix elements squared are indicated by the different colors going from white, light grey (grey), medium grey (light blue), grey (dark blue) and dark grey (red) for $\tilde{B},\tilde{W}^0,\tilde{H}_1^0,\tilde{H}_2^0$ and $\tilde{S}$, respectively. The neutralino masses are given by the numbers in GeV below the bars of the mixing content. Within the CMSSM the lightest neutralino is almost a pure bino and heavy in contrast to the light, singlino-like WIMP in the NMSSM. 
}\label{f2}
 \end{figure} 

Since we use GUT scale input parameters, the mass spectrum at the low mass SUSY scales is calculated via the renormalization group equations (RGEs), so the masses are correlated. The gaugino masses are proportional to $m_{1/2}$ \cite{Martin:1997ns,Haber:1984rc,deBoer:1994dg}:
\beq\label{eq2}
M_1\approx 0.4 m_{1/2},~ 
M_2\approx 0.8 m_{1/2},~
M_3\approx M_{\tilde{g}} \approx 2.7 m_{1/2}.  
\eeq 

Here we have included $M_3$, the gluino mass parameter of the $SU(3)$ group. In the CMSSM the value of the Higgs mixing parameter $\mu$ is given by electroweak symmetry breaking, which leads to $\mu > m_{1/2}$, so typically $M_1 < \mu$, which implies from  Eq. \ref{eq1} that in the CMSSM the lightest neutralino is usually  bino-like with a mass approximately 0.4$m_{1/2}$. The 126 GeV Higgs mass requires heavy stop masses, which can be obtained for large values of \mzero~and/or \mhalf. Scenarios with large \mzero~and low \mhalf~are excluded by the direct dark matter searches \cite{Beskidt:2012sk}, so the combination of the Higgs mass and direct searches leads to a lower limit on \mhalf, which leads to a lower limit of 180 GeV on the lightest neutralino in the CMSSM. The neutralino mass eigenstates are obtained from the diagonalization of ${\cal M}_0$ in Eq. \ref{eq1} and are linear combinations of the gaugino and Higgsino states: 
\beq
\footnotesize{
\chi^0_i = \tilde{\cal M}_0(i,1) \left | \tilde{B} \right\rangle +\tilde{\cal M}_0(i,2) \left | \tilde{W}^0 \right\rangle 
  +\tilde{\cal M}_0(i,3) \left | \tilde{H}^0_1 \right\rangle +\tilde{\cal M}_0(i,4) \left | \tilde{H}^0_2 \right\rangle+\tilde{\cal M}_0(i,5) \left | \tilde{S} \right\rangle.}
\eeq
The coefficients $\tilde{\cal M}_0(i,j)^2$ are plotted in Fig. \ref{f2a} for each of the four CMSSM neutralinos and in Fig. \ref{f2b} for the five neutralinos of the NMSSM for typical mass points\footnote{In Fig. \ref{f2a} we use: \mzero=2500 GeV, \mhalf=2375 GeV, $A_0$=-4999 GeV, \tb=48.1, sign $\mu>$0. In Fig. \ref{f2b} we use: $m_0$=2450 GeV, $m_{1/2}$=550 GeV, $A_0$=-1842 GeV, \tb=4.17, $A_{\kappa}$=2486 GeV, $A_{\lambda}$=1754 GeV, $\kappa$=0.09, $\lambda$=0.68, $\mu_{eff}$=229 GeV.}.

One observes that in the CMSSM (NMSSM) the lightest neutralino is largely a bino (singlino). We first discuss the NMSSM case. The vacuum expectation value of the Higgs singlet $\langle s \rangle$ is usually taken to be of the order of the electroweak scale:
\beq\label{eq3}
\mu_{eff}=\lambda \langle s \rangle,
\eeq
where $\lambda$ is the coupling of the Higgs doublets with the singlet. Since the vev of the singlet $\langle s \rangle$ is typically of the order of the electroweak scale, the element ${\cal M}(5,5)$ in Eq. \ref{eq1} is the lightest element. In this case the lightest neutralino is singlino-like with a mass independent of $m_{1/2}$, so neither the mass limits from the LHC SUSY searches nor the Higgs mass affect the WIMP mass.

 \begin{figure}[t!]
\hspace{0.08\textwidth}\textbf{\textsf{CMSSM \hspace{0.35\textwidth} NMSSM}}  
\centering
\subfloat[]{\label{f3a} \includegraphics[width=0.49\textwidth]{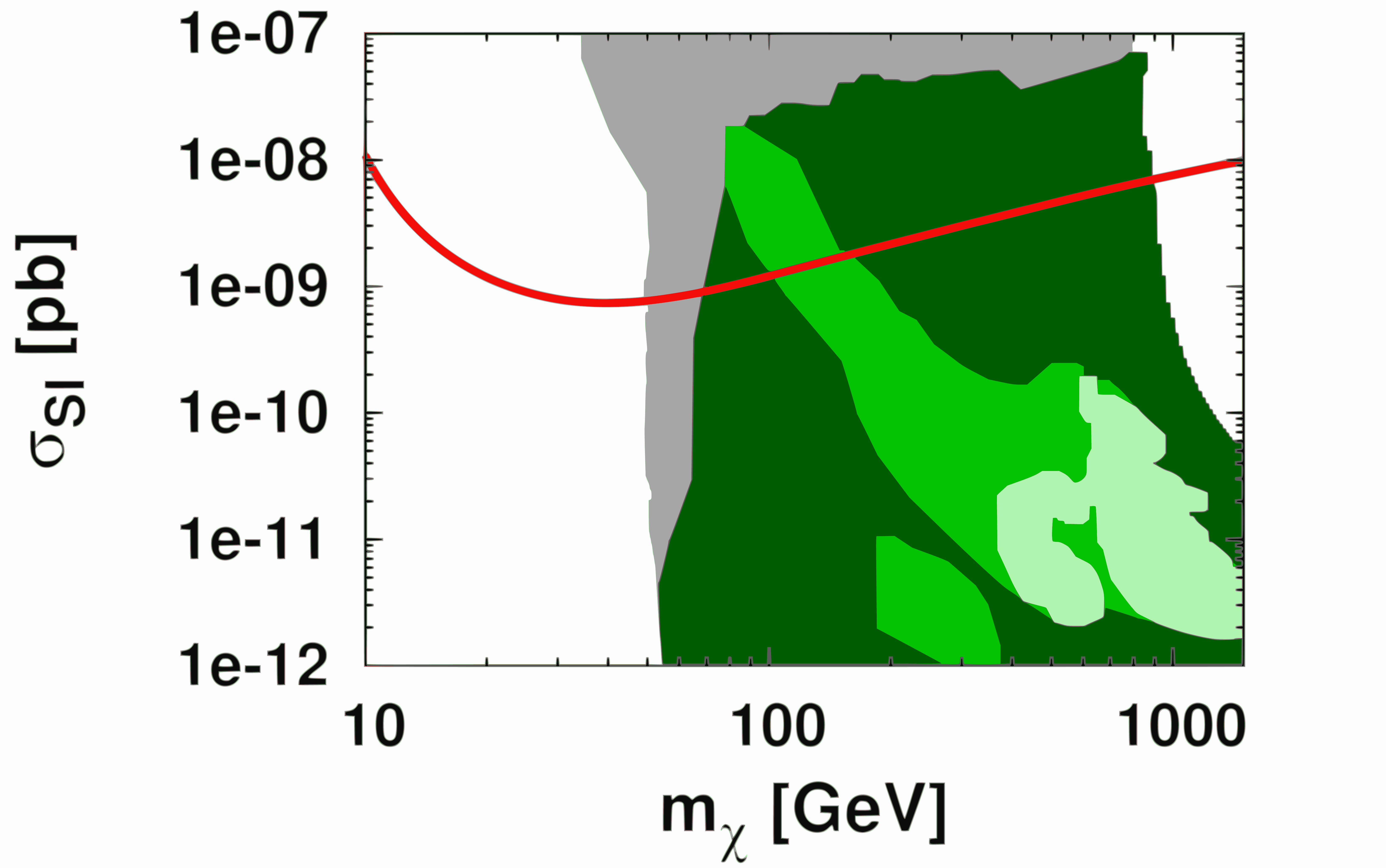} }
\subfloat[]{\label{f3b} \includegraphics[width=0.49\textwidth]{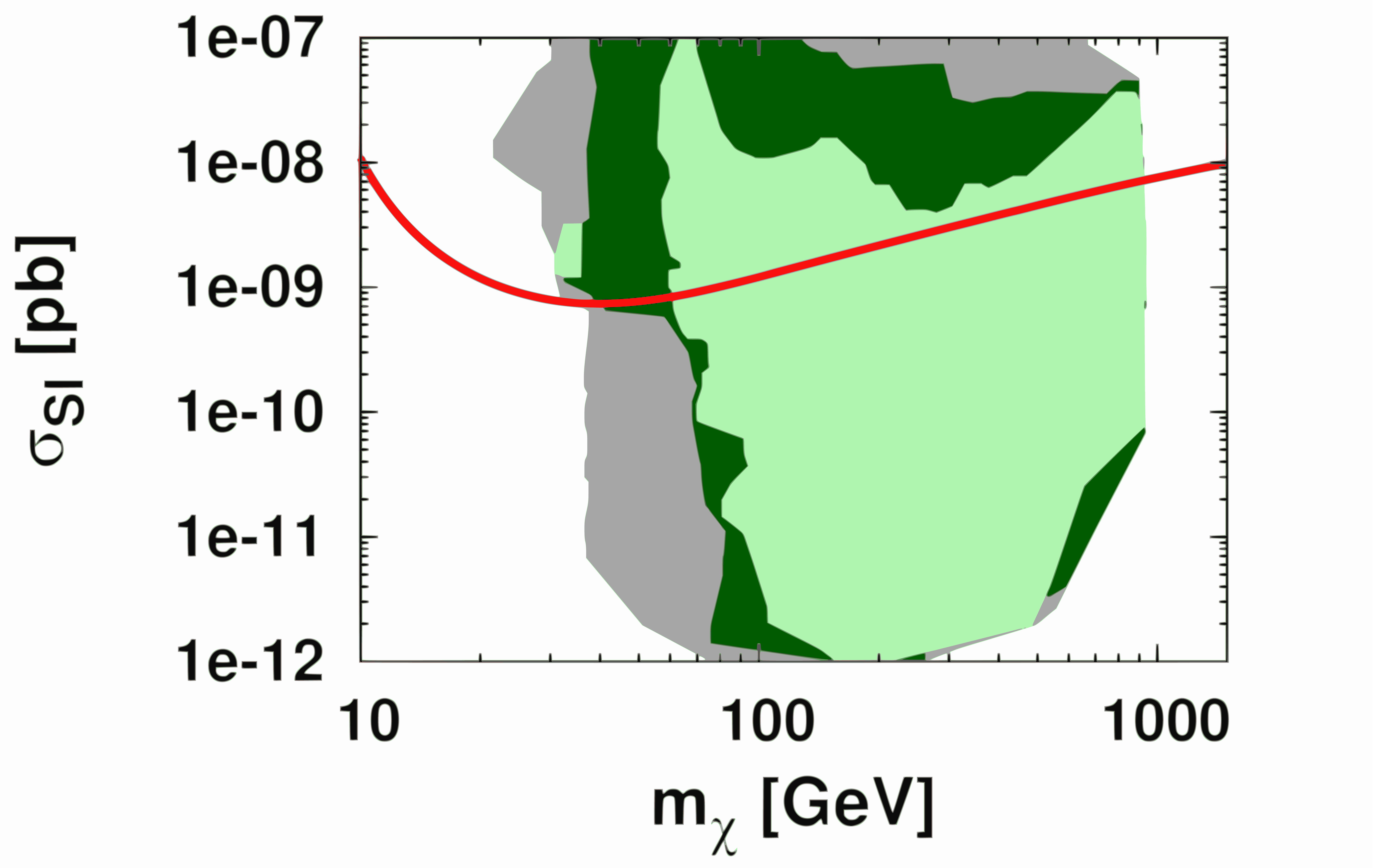} }\\
 \includegraphics[width=0.9\textwidth]{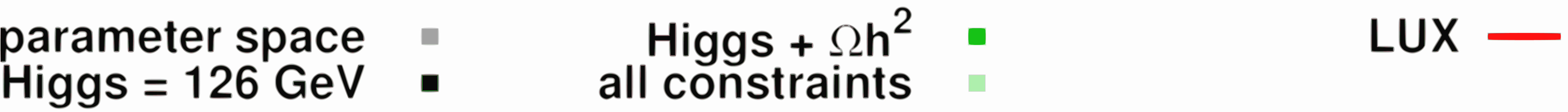} 
 \caption{ The allowed region in the (SI) WIMP-nucleon cross section versus WIMP mass plane after different constraints, as indicated in the legend below the figures, for the CMSSM (a) and NMSSM (b), respectively. All constraints means the constraints 1-9 in Table \ref{t1}. The LUX constraint No. 10 is indicated by the solid (red) line. Since in the NMSSM the WIMP mass is independent of the SUSY masses, the allowed regions are shown for a fixed mass point for (b), while the regions in (a) include a scan over all mass points.
}\label{f3}
 \end{figure}

Since we do not have constraints on $\langle s \rangle$, one can choose it to be heavy as well. If chosen above $M_1$ in the mass matrix of Eq. \ref{eq1} the lightest neutralino is not the singlino anymore, but it becomes bino-like, like in the CMSSM, as shown in Fig. \ref{f2a}. However, this is only allowed in the semi-constrained NMSSM in a very restricted region of parameter space, namely if the lightest Higgs has SM-like couplings. In most cases the second-lightest Higgs has SM couplings. To obtain the reverse, i.e. the lightest Higgs boson has SM  couplings requires a strong fine tuning of the rather large trilinear couplings as discussed in \cite{Beskidt:2013gia}. So in practically all regions of parameter space of the NMSSM (CMSSM) the LSP is singlino-like (bino-like).

We calculated the possible spin-independent (SI) WIMP-nucleon cross section and the corresponding neutralino mass for allowed points in the parameter space, both for the CMSSM and NMSSM. The results are shown in Fig. \ref{f3} by the shaded (colored) regions. In the NMSSM the mass of the lightest neutralino is independent on \mzero~and \mhalf, so the allowed regions in Fig. \ref{f3b} are generated for \mzero=\mhalf=2000 GeV. The allowed regions are further divided into regions which have 
\begin{itemize}
\item $m_h = (126 \pm 2)$ GeV (dark (green) region)
\item $m_h$ and $\Omega h^2 = 0.1199 \pm 0.0139$  (medium (green) region)
\item  $m_h$, $\Omega h^2$ and the remaining constraints in Table \ref{t1} except the LUX limit (light (green) region)
\end{itemize}
at 95\% C.L.. As shown in Fig. \ref{f3}, a large region is already excluded by the direct dark matter searches from LUX (above the solid (red) line), which will be discussed in more detail in section \ref{xenon}. 

In the CMSSM the WIMP mass can reach large values, since $m_{WIMP} \propto \mhalf$. The lower limit on the WIMP mass in the CMSSM is given by the LEP limit on the chargino mass \cite{Abbiendi:2003sc}. The Higgs constraint reduces the allowed parameter space only slightly as shown by the dark (green) region. Adding the dark matter constraint narrows the allowed range of $\sigma_{SI}$ (medium green). Adding all other constraints requires in addition heavy SUSY masses, mainly because \bsmm~needs a suppression by heavy SUSY masses, as discussed before, which leads to a lower limit on the neutralino mass of about 360 GeV.

In the NMSSM the singlino-like neutralino ranges from 20 to 1000 GeV. The lower limit results from the fact, that $\mu_{eff}$ cannot be arbitrary small, otherwise the lightest Higgs mass squared is getting negative. High WIMP values can only be obtained, if all diagonal elements in Eq. \ref{eq1} are large, which requires $2 \kappa s$ to be chosen large. The whole WIMP mass range is compatible with the combination of all constraints\footnote{Both possible Higgs scenarios are considered, i.e. either the lightest or the second lightest Higgs is allowed to be the SM Higgs boson.}. The LUX experiment limits the allowed cross section $\sigma_{SI}$ to values below $10^{-8}$pb for both models.

\subsection{WIMP-nucleon cross section in the NMSSM}\label{xenon}

\begin{figure}[t!]
 \centering
\subfloat[]{\label{f4a}  \includegraphics[width=0.45\textwidth]{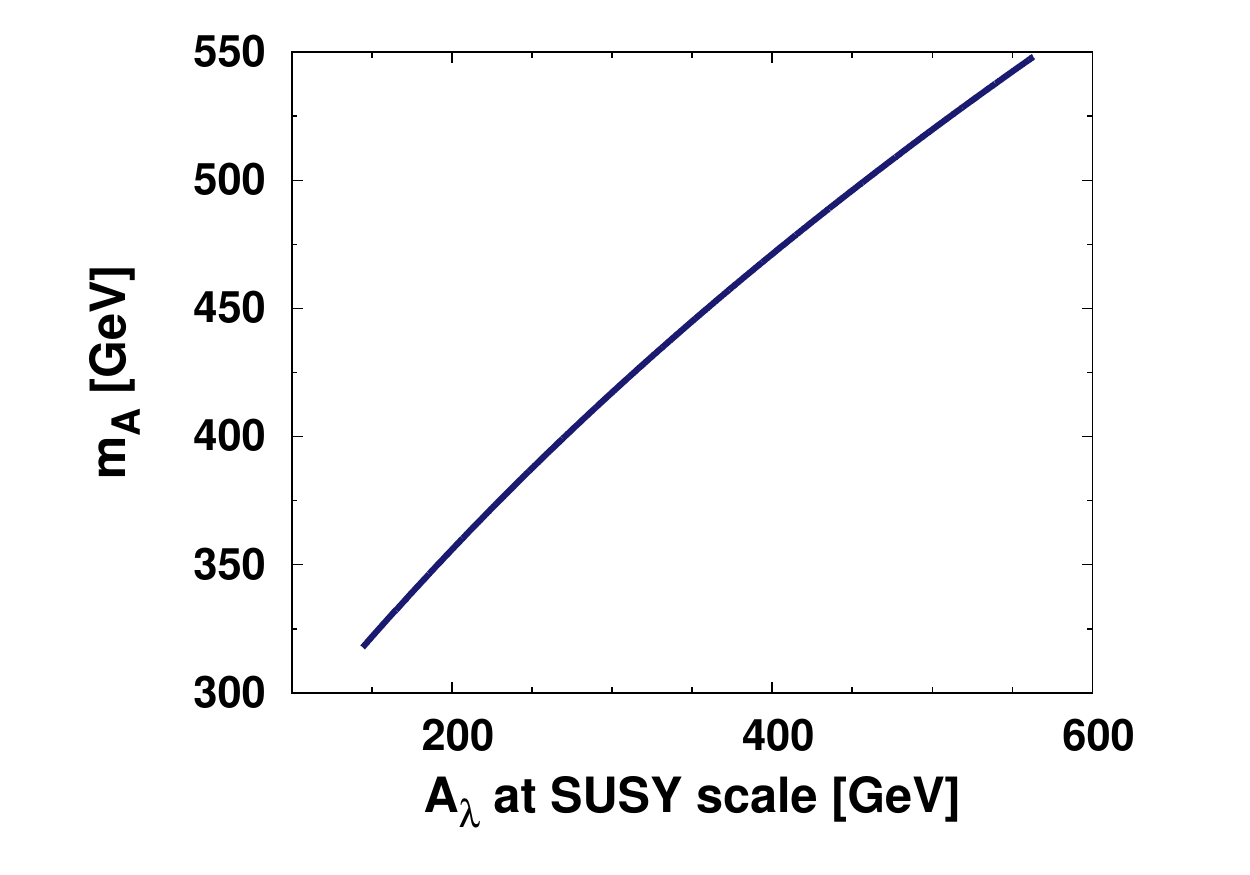}} \hspace{0.05\textwidth}
\subfloat[]{\label{f4b}  \includegraphics[width=0.45\textwidth]{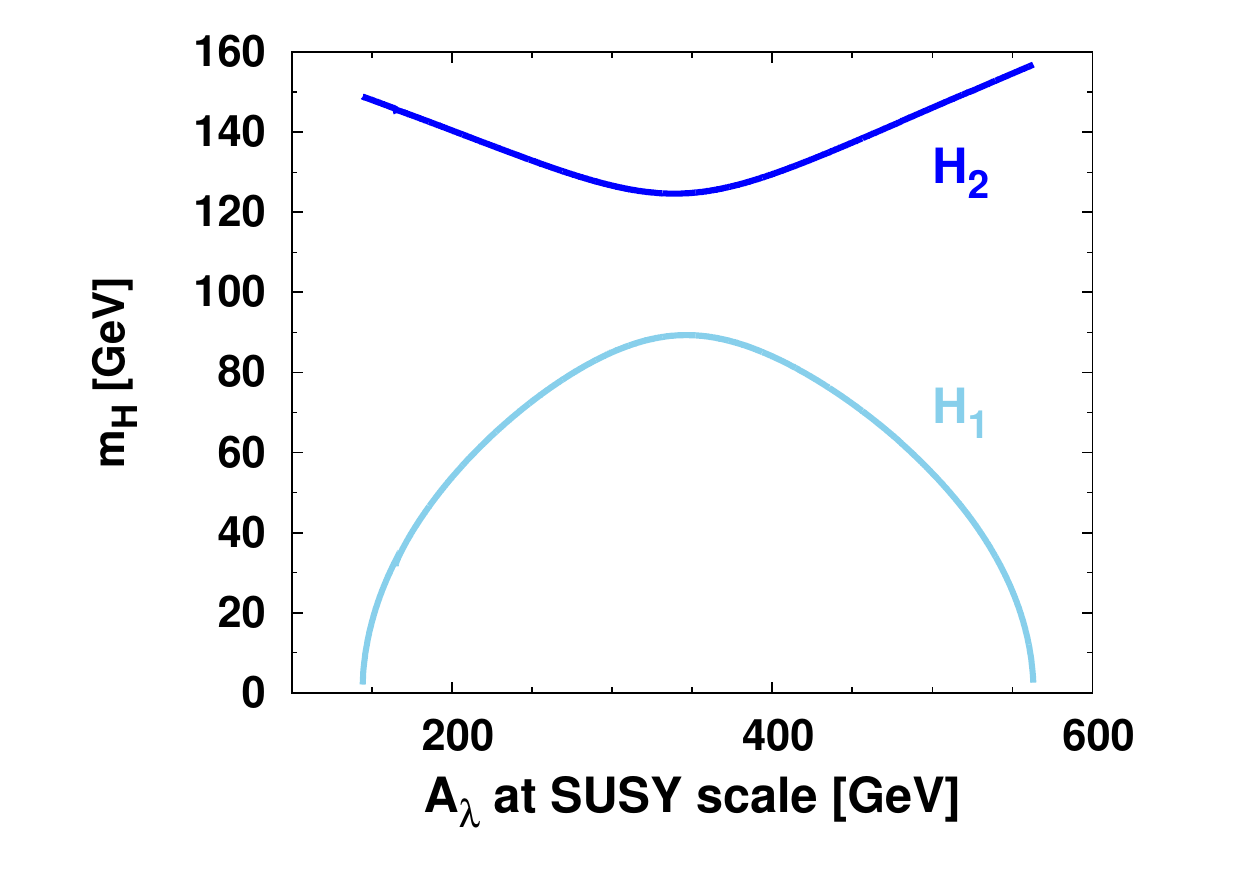}}\\
\subfloat[]{\label{f4c}  \includegraphics[width=0.45\textwidth]{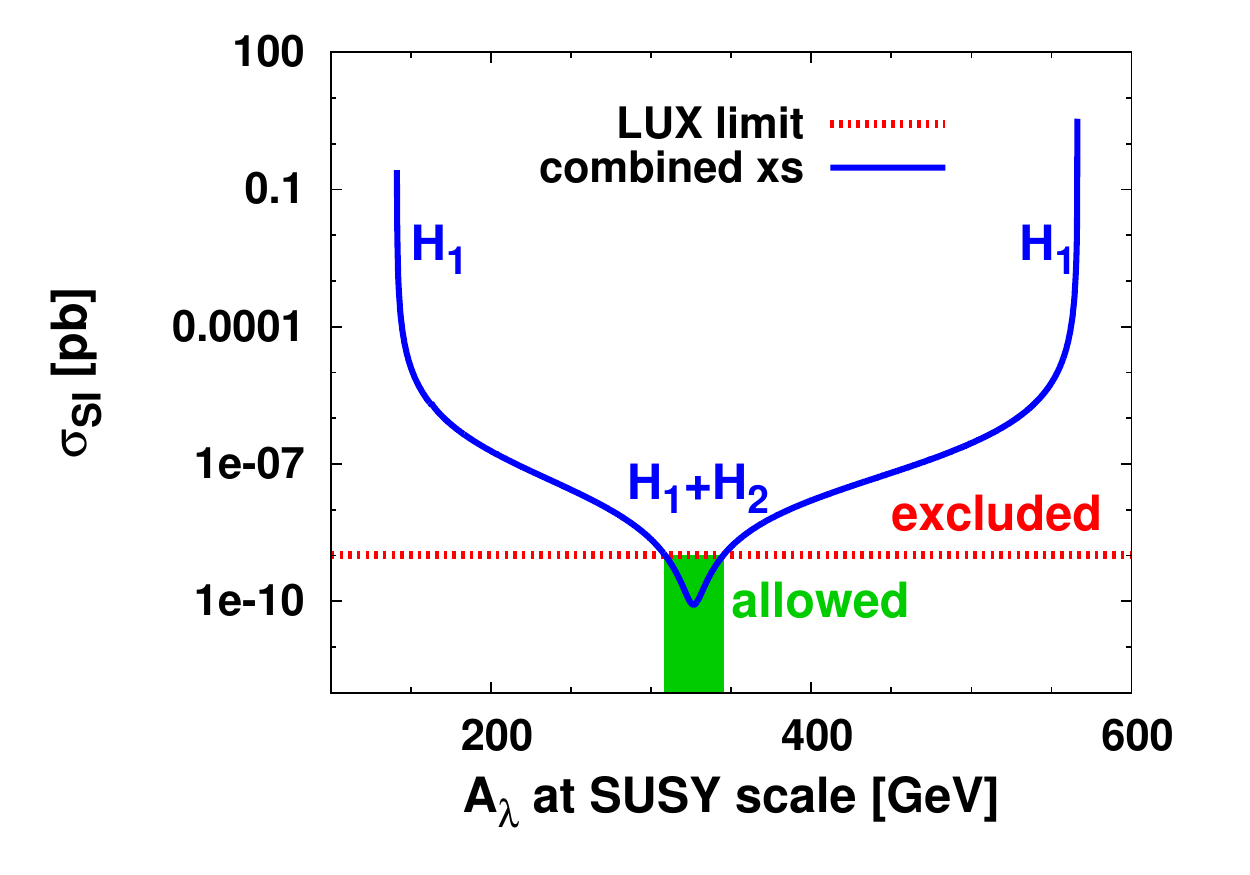}} \hspace{0.05\textwidth}
\subfloat[]{\label{f4d}  \includegraphics[width=0.45\textwidth]{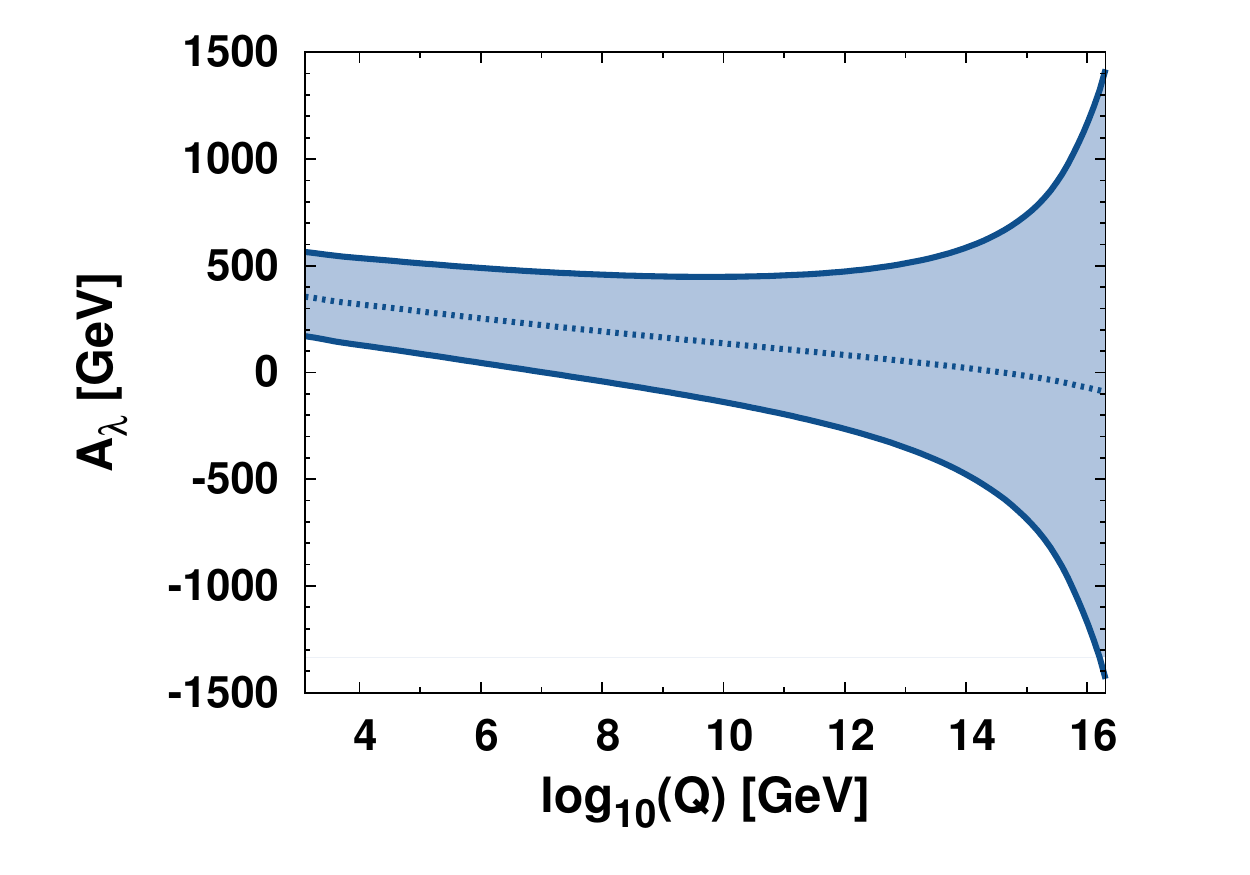} }
 \caption{ The Higgs masses $m_A$ (a) and $m_{H_{1(2)}}$ (b) are plotted as a function of $A_{\lambda}$ at the low energy SUSY scale. (c): Spin-independent WIMP-nucleon cross section for the combination of proton and neutron (solid (blue) line) plotted versus $A_{\lambda}$. The cross section has a minimum, if the masses of the two lightest Higgs bosons are similar, in which case both amplitudes are similar and almost perfectly cancel each other by the negative interference. (d): Running of $A_{\lambda}$ from the GUT scale to low scales, leading to a quasi-fixed-point solution consistent with the LUX limits, as shown in (c). Note that the range of $m_A$ values in (a) can be increased to higher values by e.g. higher values of $\mu_{eff}$.
}\label{f4}
 \end{figure}

To detect dark matter in direct DM searches one has to measure the recoil of a WIMP scatter on a nucleus. 
Several experiments try to measure these rare events, but no dark matter particle has been detected so far. The best limit for the spin-independent (SI) WIMP-nucleon cross section is given so far by the LUX experiment \cite{Akerib:2013tjd}, but other experiments give low limits as well \cite{V.Y.KozlovfortheEDELWEISS:2013mba,Aprile:2012nq,Ahmed:2009zw,Angloher:2011uu}. They exclude discovery claims by DAMA/LIBRA \cite{Bernabei:2010mq} and CoGeNT \cite{Aalseth:2012if}. The main contribution to the scalar elastic scattering amplitude of a neutralino scattering on quarks comes from scalar Higgs boson t-channel exchange. The pseudo-scalar Higgs boson exchange is suppressed because of parity, whereas the heavy squark exchange as well as the heavy Higgs boson exchange are suppressed by their mass. So the scattering via the lightest $H_1$ and the second lightest Higgs boson $H_2$ is dominant in the NMSSM. These diagrams have a negative interference, which can lead to very small cross sections, especially if the masses of $H_1$ and $H_2$ are similar. $H_1$ and $H_2$ depend both on the pseudo-scalar Higgs mass, which in turn is a function of $A_{\lambda}$ at the low energy SUSY scale. The dependence of $m_A$ and $m_{H_{1(2)}}$ on $A_\lambda$ at the SUSY scale is shown in the top row of Fig. \ref{f4}. One notices that $m_{H_1}$ can become zero for small and large values of $A_{\lambda}$ and the spin-independent cross section becomes correspondingly large, as shown in Fig. \ref{f4c}. For $A_\lambda$ values in between, the values of $m_{H_1}$ and $m_{H_2}$ become similar and $\sigma_{SI}$ becomes small. The cross section can actually become zero for either a proton or a neutron, but this does not happen for equal regions of parameter space. Therefore, the average cross section stays finite in Fig. \ref{f4c}. The horizontal dotted (red) line in Fig. \ref{f4c} corresponds to the LUX limit, which excludes a wide range of $A_{\lambda}$. The allowed values of $A_{\lambda}$ are exactly in the range of the quasi-fixed point solutions of the RGEs, as shown in Fig. \ref{f4d}.

\subsection{Future prospects}
\label{future}
\begin{figure}[t!]
\hspace{0.08\textwidth}\textbf{\textsf{CMSSM \hspace{0.35\textwidth} NMSSM}}  
 \centering
\subfloat[]{\label{f5a}   \includegraphics[width=0.49\textwidth]{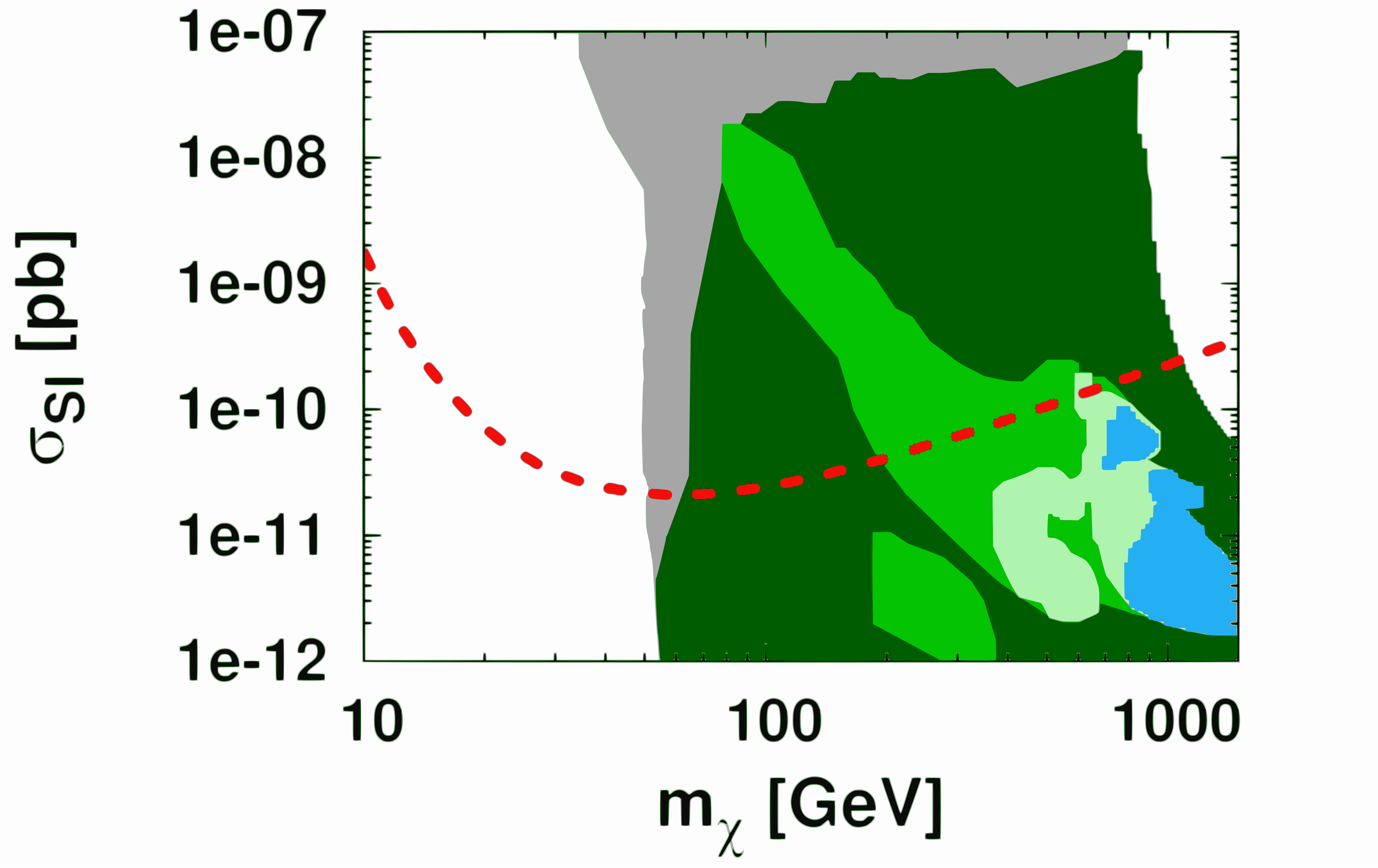}  }
\subfloat[]{\label{f5b}   \includegraphics[width=0.49\textwidth]{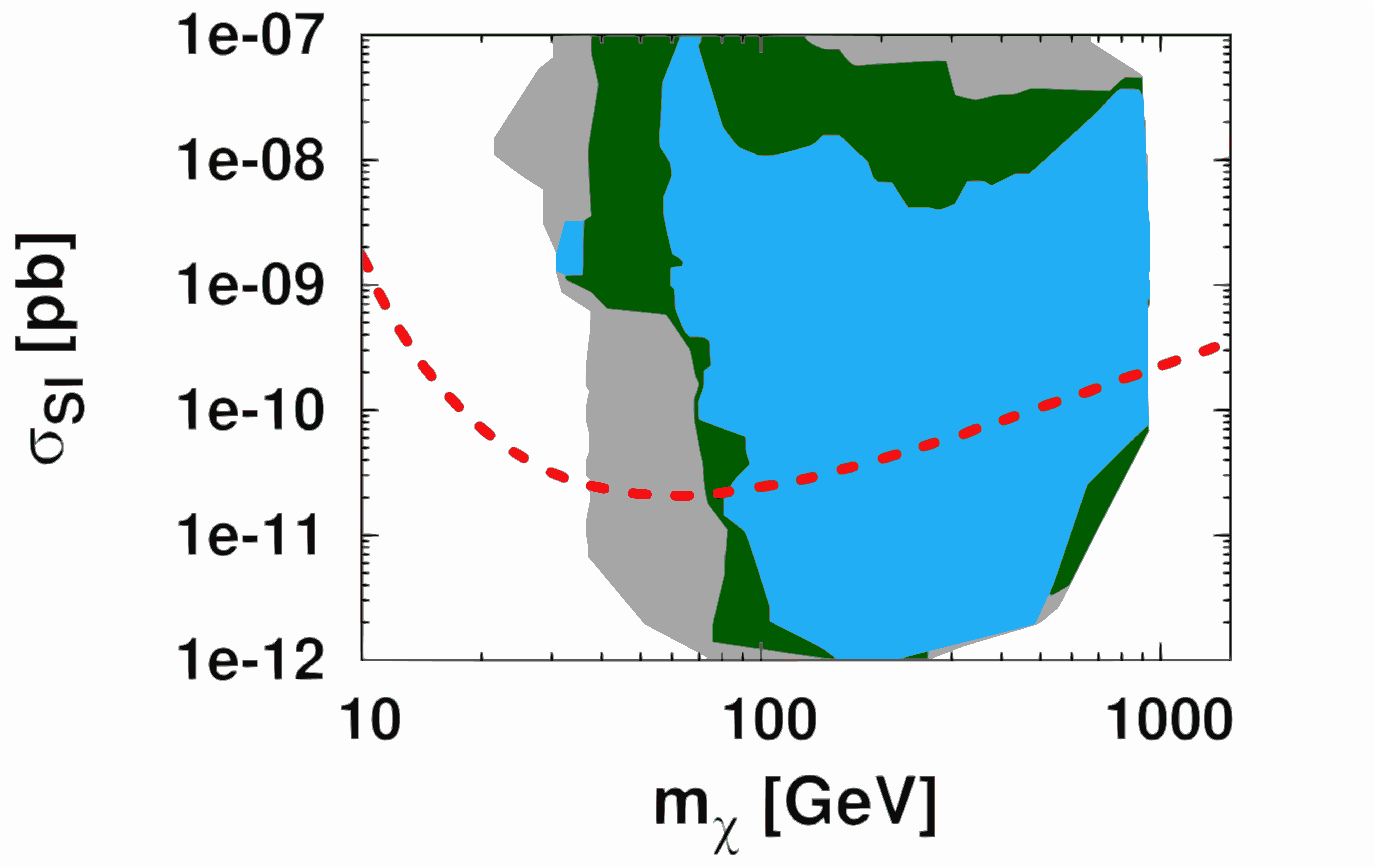}}\\
   \includegraphics[width=0.9\textwidth]{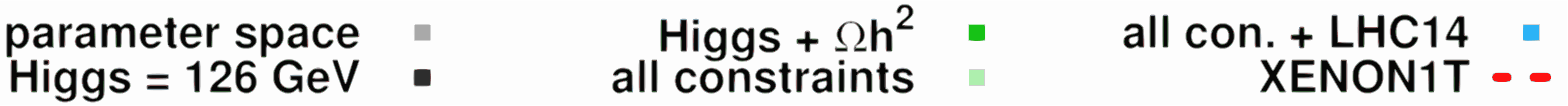}
 \caption{ Same optimized regions as in Fig. \ref{f3} for the CMSSM (a) and NMSSM (b) including expected sensitivities for future searches. The regions above the dashed (red) lines are sensitive to XENON1T. The light (blue) central regions will not be accessible to future LHC SUSY searches at 14 TeV and 3000$fb^{-1}$ (LHC14).  
}\label{f5}
 \end{figure}

Future searches are needed to probe the remaining allowed parameter space, since no SUSY or dark matter particles have been found so far. We show the prospects for discovery by extrapolating the current sensitivities for future direct dark matter experiments and LHC at 14 TeV. 
To determine the sensitivity for future direct dark matter experiments we parameterize the limit given by XENON100 and extrapolate to the XENON1T limit, which is expected to reach a sensitivity two orders of magnitude better than XENON100 \cite{Aprile:2012nq}. 
For the future prospects of an energy of 14 TeV  at the LHC with an integrated luminosity of 3000$fb^{-1}$ we extrapolate the current cross section limits of the SUSY searches at the LHC for squarks and gluinos, which exclude low SUSY masses and accordingly large cross sections. The hadronic searches are the most sensitive ones, so  the 95\% C.L. exclusion contours in the \mzero-\mhalf~plane \cite{ATLAS-CONF-2013-047,ATLAS-CONF-2013-061} determine the limits on the hadronic cross sections $\sigma_{tot}(pp\rightarrow \tilde{g}\tilde{g},\tilde{g}\tilde{q},\tilde{q}\tilde{q})$, which vary along the contour because of the varying efficiency. Using Wilks's theorem one can show that the profile likelihood ratio leads to a $\chi^2$-distribution for the hadronic cross section: 
\beq\label{eq4}
\chi^2=\left (\frac{\sigma_{95}-\sigma_0}{error} \right )^2\approx a \cdot \sigma_{tot}^2=\frac{a N^2}{\epsilon^2 L^2}, 
\eeq

where $N$ represents the total number of events with the corresponding efficiency $\epsilon$ and integrated luminosity $L$.  
The proportionality factor $a$ can be determined as a function of \mzero~by the requirement of a 95\% C.L. exclusion on the contour line: $\Delta \chi^2=5.99=a\cdot \sigma_{tot}^2$. To obtain the limit on the cross section at a higher luminosity $L$ we scale the limit with 1/$L^2$ and then check for each point in the \mzero~and \mhalf~plane, where the limit is reached for 14 TeV. This method was tested to work, if we extrapolate from the early  low luminosity results at 7 TeV to the high luminosity results at 8 TeV.   
The projected exclusion contour at 14 TeV is shown by the  dotted (red) lines in the top right corners of Fig. \ref{f1}. Clearly, the LHC running at 14 TeV will be sensitive to squarks and gluinos around 3.5 and 3.1 TeV, respectively. To compare this sensitivity with the direct searches we repeat in Fig. \ref{f5} the plot from Fig. \ref{f3} and add as a dashed (red) line the expected limits from XENON1T. The regions above these contours are excluded. In addition, the light (blue) central area corresponds to the region, which will  escape the sensitivity from LHC searches at 14 TeV. In the CMSSM the non-accessible region occurs only for large WIMP masses, since the LHC SUSY searches are only sensitive to gluino masses up to 3.1 TeV (Fig. \ref{f1}). This implies sensitivities to WIMPs around 600 GeV, since $m_{\tilde{g}}/m_{WIMP} \approx M_3 / M_1 \approx 5-6.75$ in the CMSSM. If the LHC SUSY searches are combined with all constraints from Table \ref{t1} the lower limit increases up to 680 GeV, see Fig. \ref{f5a}. In the NMSSM such a relation to the gluino mass for a singlino-like WIMP does not exist and light singlino-like WIMPs can only be probed efficiently by the direct dark matter searches. Light WIMPs, as claimed in Ref. \cite{Daylan:2014rsa}, would only be allowed in the NMSSM and exclude the CMSSM. Searches for DM particles at the LHC in monojet events \cite{CMS-PAS-EXO-12-048} and in searches for invisible decays of Higgs bosons \cite{Chatrchyan:2014tja} can currently cover WIMP-nucleon cross section down to $10^{-8}$ pb. However, the limits including the extrapolated sensitivities will still be above the limit reached by the direct dark matter searches, so we did not include these searches into the LHC searches.

\section{Summary}
In this Letter we compared the dark matter sector of the CMSSM and the NMSSM using GUT scale input parameters. Within the CMSSM the lightest neutralino is bino-like in a large region of parameter space and has a mass proportional to $\mhalf$. As shown in Fig. \ref{f3a} the Higgs mass of 126 GeV allows all neutralino masses above the LEP limit of 55 GeV in the CMSSM, but if the Higgs mass constraint is combined with the relic density, only a rather narrow stripe is allowed. If all other constraints from Table \ref{t1} are included, only the tail of this stripe is allowed, which corresponds to neutralino masses above 360 GeV.
The lightest neutralino in the NMSSM is typically singlino-like, see Fig. \ref{f2b}. The Higgs self-interactions are required to be large enough to have an annihilation cross section consistent with the relic density, but this leads typically to a WIMP-nucleon cross section above the present experimental limits (No. 10 in Table \ref{t1}). However, this cross section has two main contributions, namely the t-channel exchange of the two lightest Higgs bosons. These interfere negatively, so if they are of the same order of magnitude, the WIMP-nucleon cross section can become small, as shown in Fig. \ref{f4c}. 
The allowed mass range for the neutralino within the NMSSM covers the region from 30 GeV upwards. This range is only mildly dependent on the Higgs mass constraint of 126 GeV and all other constraints, as shown in Figs. \ref{f3b} and \ref{f5b}.
The LHC SUSY searches at 14 TeV will be able to access the multi-TeV range of squarks and gluinos and consequently will be sensitive to WIMP masses up to 680 GeV in the CMSSM if the searches are combined with all other constraints in Table \ref{t1} as shown by the light (blue) central region in Fig. \ref{f5a}. However, these searches hardly limit the singlino-like WIMP masses in the NMSSM, as shown by the light (blue) central region in Fig. \ref{f5b}. For both models the spin independent cross section down to $~10^{-11}$pb will be probed by future direct DM searches, as shown in Fig. \ref{f5}.

\section*{Acknowledgements}
Support from the Heisenberg-Landau program and the Deutsche Forschungsgemeinschaft (DFG) via a Mercator Professorship (Grant 42/577-1-560249 for Prof. Kazakov) is warmly acknowledged.
We thank U. Ellwanger for helpful discussions regarding NMSSMTools. 






\bibliographystyle{lucas_unsrt}
\input{dark_matter.bbl}


\end{document}

%% file: config.tex
\newlength{\dslashwidth}

\newcommand{\bsg}{\ensuremath{b\to X_s\gamma}}

\newcommand{\tb}{\ensuremath{\tan\beta}}

\newcommand{\beq}{\begin{equation}} 
\newcommand{\eeq}{\end{equation}}
\newcommand{\beqa}{\begin{eqnarray}} 
\newcommand{\eeqa}{\end{eqnarray}}
\newcommand{\newc}{\newcommand}
\newcommand{\bq}{\begin{equation}}
\newcommand{\eq}{\end{equation}}
\newcommand{\ba}{\begin{array}}
\newcommand{\ea}{\end{array}}
\newcommand{\bqa}{\begin{eqnarray}}
\newcommand{\eqa}{\end{eqnarray}}

\newcommand{\lnf}{{\ifmmode \Lambda^{(N_f)} \else $\Lambda^{(N_f)}$\fi}}
\newcommand{\ms}{{\ifmmode \overline{MS} \else $\overline{MS}$\fi}}
\newcommand{\dr}{{\ifmmode \overline{DR} \else $\overline{DR}$\fi}}
\newcommand{\lms}{{\ifmmode \Lambda^{(5)}_{\overline{MS}} \else $\Lambda^{(5)}_{\overline{MS}}$\fi}}
\newcommand{\lam}{{\ifmmode \Lambda \else $\Lambda$\fi}}
\newcommand{\mev}{{\ifmmode {\rm MeV} \else ${\rm MeV}$\fi}}
\newcommand{\gev}{{\ifmmode {\rm GeV} \else ${\rm GeV}$\fi}}
\newcommand{\gevc}{{\ifmmode {\rm GeV/c^2} \else ${\rm GeV/c^2}$\fi}}
\newcommand{\tev}{{\ifmmode {\rm TeV} \else ${\rm TeV}$\fi}}
\newcommand{\tevc}{{\ifmmode {\rm TeV/c^2} \else ${\rm TeV/c^2}$\fi}}
\newcommand{\lp}{{\ifmmode L^+  \else $L^+$\fi}}
\newcommand{\lm}{{\ifmmode L^-  \else $L^-$\fi}}
\newcommand{\mlp}{{\ifmmode M(L^-) \else $M(L^-)$\fi}}
\newcommand{\mlz}{{\ifmmode M(L^0) \else $M(L^0)$\fi}}
\newcommand{\lz}{{\ifmmode L^0 \else $L^0$\fi}}
\newcommand{\ev}{{\ifmmode GeV/c^2 \else $GeV/c^2$\fi}}
\newcommand{\tri}{{\ifmmode \triangleup \else $\triangleup$\fi}}
\newcommand{\unl}{{\ifmmode U_{lL^0} \else $U_{lL^0}$\fi}}\newcommand{\gL}{{\ifmmode g_L \else $g_{L}$\fi}}
\newcommand{\gR}{{\ifmmode g_R  \else $g_{R}$\fi}}
\newcommand{\gumu}{{\ifmmode \gamma^{\mu} \else $\gamma^{\mu}$\fi}}
\newcommand{\gunu}{{\ifmmode \gamma^{\nu} \else $\gamma^{\nu}$\fi}}
\newcommand{\gdmu}{{\ifmmode \gamma_{\mu} \else $\gamma_{\mu}$\fi}}
\newcommand{\gdnu}{{\ifmmode \gamma_{\nu} \else $\gamma_{\nu}$\fi}}
\newcommand{\stw}{{\ifmmode\sin^2\theta_W \else $\sin^{2}\theta_{W}$ \fi}}
\newcommand{\sws}{{\ifmmode \;\sin^2\theta_W  \else $\;\sin^{2}\theta_{W}$ \fi}}
\newcommand{\cws}{{\ifmmode \;\cos^2\theta_W  \else $\;\cos^{2}\theta_{W}$ \fi}}
\newcommand{\sw}{{\ifmmode \;\sin\theta_W  \else $\sin\theta_{W}$ \fi}}
\newcommand{\cw}{{\ifmmode \;\cos\theta_W  \else $\;\cos\theta_{W}$ \fi}}
\newcommand{\tw}{{\ifmmode \;\tan\theta_W  \else $\;\tan\theta_{W}$ \fi}}
\newcommand{\qq}{{\ifmmode q\overline{q} \else $q\overline{q}$\fi}}
\newcommand{\lR}{{\ifmmode l_R \else $l_R$\fi}}
\newcommand{\lL}{{\ifmmode l_L \else $l_L$\fi}}
\newcommand{\nt}{{\ifmmode \nu_{\tau} \else $\nu_{\tau}$\fi}}
\newcommand{\nuR}{{\ifmmode \nu_R  \else $\nu_R$\fi}}
\newcommand{\nuL}{{\ifmmode \nu_L  \else $\nu_L$\fi}}
\newcommand{\qR}{{\ifmmode g_R  \else $q_R$\fi}}
\newcommand{\qL}{{\ifmmode q_L  \else $q_L$\fi}}
\newcommand{\qRp}{{\ifmmode q_R'  \else $q_{R}$'\fi}}
\newcommand{\qLp}{{\ifmmode q_L'  \else $q_{L}$'\fi}}
\newcommand{\est}{{\ifmmode e^{\bf \ast} \else $e^{\bf \ast}$\fi}}
\newcommand{\lst}{{\ifmmode l^{\bf \ast} \else $l^{\bf \ast}$\fi}}
\newcommand{\must}{{\ifmmode \mu^{\bf \ast} \else $\mu^{\bf \ast}$\fi}}
\newcommand{\taust}{{\ifmmode \tau^{\bf \ast} \else $\tau^{\bf \ast}$ \fi}}
\newcommand{\pperp}{{\ifmmode p_t  \else $p_t$\fi}}
\newcommand{\et}{{\ifmmode E_t  \else $E_t$\fi}}
\newcommand{\xt}{{\ifmmode x_t  \else $x_t$\fi}}
\newcommand{\smumu}{{\ifmmode \sigma_{\mu\mu}  \else $\sigma_{\mu\mu}$ \fi}}
\newcommand{\eg}{{\ifmmode e\gamma  \else $e\gamma$\fi}}
\newcommand{\epem}{{\ifmmode e^+e^-  \else $e^+e^-$\fi}}
\newcommand{\lplm}{{\ifmmode L^+L^-  \else $L^+L^-$\fi}}
\newcommand{\pp}{{\ifmmode p\overline p  \else $p\overline p$\fi}}
\newcommand{\llz}{{\ifmmode L^0\overline{L}^0 \else $L^0\overline{L}^0$\fi}}
\newcommand{\epemt}{{\ifmmode e^+e^- \to  \else $e^+e^- \to$\fi}}
\newcommand{\eb}{{\ifmmode E_{beam}  \else $E_{beam}$\fi}}
\newcommand{\ip}{{\ifmmode pb^{-1}  \else $pb^{-1}$\fi}}
\newcommand{\upm}{{\ifmmode ^{\pm}  \else $^{\pm}$\fi}}
\newcommand{\de}{{\ifmmode ^{\circ}  \else $^{\circ}$ \fi}}
\newcommand{\appr}{{\ifmmode \sim \else $\sim$ \fi}}
\newcommand{\corresp}{{\ifmmode \stackrel{\wedge}{=} \else $\stackrel{\wedge}{=}$ \fi}}
\newcommand{\sqrts}{{\ifmmode \sqrt{s} \else $\sqrt{s}$\fi}}
\newcommand{\zz}{{\ifmmode Z^0  \else $Z^0$\fi}}
\newcommand{\mz}{{\ifmmode M_{Z}  \else $M_{Z}$\fi}}
\newcommand{\mzs}{{\ifmmode M_{Z}^2  \else $M_{Z}^2$\fi}}
\newcommand{\mw}{{\ifmmode M_{W}  \else $M_{W}$\fi}}
\newcommand{\mws}{{\ifmmode M_{W}^2  \else $M_{W}^2$\fi}}
\newcommand{\mh}{{\ifmmode M_{Higgs}  \else $M_{Higgs}$\fi}}
\newcommand{\gt}{{\ifmmode \Gamma_{tot} \else $\Gamma_{tot}$\fi}}
\newcommand{\msusy}{{\ifmmode M_{SUSY}  \else $M_{SUSY}$\fi}}
\newcommand{\msusys}{{\ifmmode M_{SUSY}^2  \else $M_{SUSY}^2$\fi}}
\newcommand{\su}{{\ifmmode SU(3)_C\otimes\- SU(2)_L\otimes\- U(1)_Y \else $SU(3)_C\otimes SU(2)_L\otimes U(1)_Y$\fi}}
\newcommand{\suthree}{{\ifmmode SU(3)_C  \else $SU(3)_C$\fi}}
\newcommand{\sutwo}{{\ifmmode  SU(2)_L\otimes U(1)_Y \else $SU(2)_L\otimes U(1)_Y$\fi}}
\newcommand{\taup}{{\ifmmode \tau_{proton} \else $\tau_{proton}$\fi}}
\newcommand{\as}{{\ifmmode \alpha_{s}  \else $\alpha_{s}$\fi}}
\newcommand{\mgut}{{\ifmmode M_{GUT}  \else $M_{GUT}$\fi}}
\newcommand{\mguts}{{\ifmmode M_{GUT}^2  \else $M_{GUT}^2$\fi}}
\newcommand{\mzero}{{\ifmmode m_0        \else $m_0$\fi}}
\newcommand{\mhalf}{{\ifmmode m_{1/2}    \else $m_{1/2}$\fi}}
\newcommand{\sq}{{\ifmmode \tilde{q}    \else $\tilde{q}$\fi}}
\newcommand{\gl}{{\ifmmode \tilde{g}    \else $\tilde{g}$\fi}}
\newcommand{\mb}{{\ifmmode m_{b}    \else $m_{b}$\fi}}
\newcommand{\mt}{{\ifmmode m_{t}    \else $m_{t}$\fi}}
\newcommand{\mts}{{\ifmmode m_{t}^2    \else $m_{t}^2$\fi}}

\newcommand{\mtau}{{\ifmmode m_{\tau}  \else $m_{\tau}$\fi}}
\newcommand{\dpp}{{\ifmmode \delta_{pert} \else $\delta_{pert}$\fi}}
\newcommand{\dnp}{{\ifmmode\delta_{non-pert}\else$\delta_{non-pert}$\fi}}
\newcommand{\dew}{{\ifmmode \delta_{\rm EW}\else $\delta_{\rm EW}$\fi}}
\newcommand{\rt}{{\ifmmode R_{\tau}  \else $R_{\tau} $\fi}}
\newcommand{\rz}{{\ifmmode R_{Z}  \else $R_{Z} $\fi}}

\newcommand{\swb}{{\ifmmode \sin^2\theta_{\overline{MS}} \else $\sin^2\theta_{\overline{MS}}$\fi}}
\newcommand{\cwb}{{\ifmmode \cos^2\theta_{\overline{MS}} \else $\cos^2\theta_{\overline{MS}}$\fi}}

\newcommand{\bsmm}{\ensuremath{B^0_s\to\mu^+\mu^-}}

\newcommand{\btaunu}{\ensuremath{B_u\to\tau\nu}}

\newc\AIPCP[3] {{\em AIP Conf. Proc.} {\bf #1} (#2) #3}
\newc\AJ[3] {{\em Astrophys. J.} {\bf #1} (#2) #3}
\newc\AMS[3] {{\em Ann. Math. Statist.} {\bf #1} (#2) #3}
\newc\AP[3] {{\em Ann. Phys.} {\bf #1} (#2) #3}
\newc\APJ[3] {{\em Astropart. J.} {\bf #1} (#2) #3}
\newc\APP[3] {{\em Astropart. Phys.} {\bf #1} (#2) #3}
\newc\APS[3] {{\em Astrophys. J. Suppl.} {\bf #1} (#2) #3}
\newc\ARNPS[3] {{\em Ann. Rev. Nucl. Part. Sci.} {\bf C#1} (#2) #3}
\newc\BA[3] {{\em Bayesian Anal.} {\bf C#1} (#2) #3}
\newc\CPC[3] {{\em Comput. Phys. Commun.} {\bf C#1} (#2) #3}
\newc\CP[3] {{\em Contemp. Phys.} {\bf #1} (#2) #3}
\newc\EPJ[3] {{\em Euro. Phys. Journ.} {\bf C#1} (#2) #3}
\newc\JCAP[3] {{\em JCAP} {\bf #1} (#2) #3}
\newc\JHEP[3] {{\em JHEP} {\bf #1} (#2) #3}
\newc\JPG[3] {{\em J. Phys.} {\bf G #1} (#2) #3}
\newc\IJMP[3] {{\em Int. J. Mod. Phys.} {\bf A #1} (#2) #3}
\newc\MNRAS[3] {{\em Mon. Not. Roy. Astron. Soc.} {\bf #1} (#2) #3}
\newc\MPL[3] {{\em Mod. Phys. Lett.} {\bf A #1} (#2) #3}
\newc\NAR[3] {{\em New Astron. Rev.} {\bf #1} (#2) #3}
\newc\NCA[3] {{\em Nuovo Cimento} {\bf #1} (#2) #3}
\newc\NIM[3] {{\em Nucl. Instrum. Methods} {\bf #1} (#2) #3}
\newc\NIMA[3] {{\em Nucl. Instrum. Methods} {\bf A #1} (#2) #3}
\newc\NAT[3] {{\em Nature} {\bf #1} (#2) #3}
\newc\NPB[3] {{\em Nucl. Phys.} {\bf B #1} (#2) #3}
\newc\NPA[3] {{\em Nucl. Phys.} {\bf A #1} (#2) #3}
\newc\NPPS[3] {{\em Nucl. Phys. Proc. Suppl.} {\bf #1} (#2) #3}
\newc\PLB[3] {{\em Phys. Lett.} {\bf B #1} (#2) #3}
\newc\PR[3] {{\em Phys. Rep.} {\bf #1} (#2) #3}
\newc\PRL[3] {{\em Phys. Rev. Lett.} {\bf #1} (#2) #3}
\newc\PRD[3] {{\em Phys. Rev.} {\bf D #1} (#2) #3}
\newc\PRC[3] {{\em Phys. Rev.} {\bf C #1} (#2) #3}
\newc\PTP[3] {{\em Prog. Theor. Phys.} {\bf #1} (#2) #3}
\newc\RMP[3] {{\em Rev. Mod. Phys.} {\bf #1} (#2) #3 }
\newc\RPP[3] {{\em Rept. Prog. Phys.} {\bf #1} (#2) #3 }
\newc\SC[3] {{\em Science} {\bf #1} (#2) #3 }
\newc\ZPC[3] {{\em Z. Phys.} {\bf C #1} (#2) #3}
\newc\Err[3] {{\em Erratum-ibid.} {\bf #1} (#2) #3 }

%% file: dark_matter.bbl
\providecommand{\href}[2]{#2}\begingroup\raggedright\endgroup